\documentclass[twocolumn,prl,aps,superscriptaddress,showpacs,amsmath,amssymb,floatfix]{revtex4-1}
\usepackage{color}
\usepackage{xcolor}
\usepackage{mathrsfs}
\usepackage{amsmath}
\usepackage{graphicx}
\usepackage{dcolumn}
\usepackage{bm}
\usepackage{times}
\usepackage{amssymb}
\usepackage{float}
\usepackage{array}
\usepackage{tikz}

\begin{document}

\title{ Topological superfluids with time-reversal symmetry from $s$-wave interaction in a bilayer system}
\author{Beibing Huang$^*$}
\affiliation{Department of Physics and Center for quantum Coherence, The Chinese University of Hong Kong Shatin, N.T., Hong Kong, China}
\affiliation{Department of Physics, Yancheng Institute of Technology, Yancheng, 224051, P. R. China}
\author{Pak Hong Chui$^*$}
\affiliation{Department of Physics and Center for quantum Coherence, The Chinese University of Hong Kong Shatin, N.T., Hong Kong, China}
\author{Jia Liu}
\affiliation{Department of Physics and Center for quantum Coherence, The Chinese University of Hong Kong Shatin, N.T., Hong Kong, China}
\author{Chuanwei Zhang}
\affiliation{Department of Physics, The University of Texas at Dallas, Richardson, TX 75080, USA}
\author{Ming Gong}
\affiliation{Department of Physics and Center for quantum Coherence, The Chinese University of Hong Kong Shatin, N.T., Hong Kong, China}

\date{\today}

\begin{abstract}
Topological superconducting phases with time-reversal (TR) symmetry have been widely explored in recent years.
However the involved unconventional pairings are generally implausible in realistic materials. Here we demonstrate via
detailed self-consistent calculation that these topological phases with TR symmetry in DIII and
BDI classes can be realized in a spin-orbit coupled bilayer system with only $s$-wave interaction. The bilayer freedom
enables the definition of TR symmetry between the layers even in the presence of local Zeeman fields, which we
propose to be realized using four laser beams. The gapped phase in DIII class is characterized by $\mathbb{Z}_2$,
while all the gapless phases in these two classes are characterized by nontrivial winding numbers and are also
manifested from the Majorana flat bands.  We also unveil the intimate relation between TR symmetry and mirror
symmetry due to phase locking effect between the two layers, which harbors the mirror symmetry protected topological
phases. We finally demonstrate that these phases will not be spoiled by interlayer pairings.
\end{abstract}

\maketitle

There is a major effort in realizing unpaired Majorana fermions
(MFs) in both condensed matter physics and cold atom physics, which
have important applications in topological quantum
computations\cite{Nayak}. These systems include $p$-wave
superconducting phases in Sr$_2$RuO$_4$\cite{SRO, SRO2},
Helium-3\cite{HeB} and fully spin polarized ultracold
atoms\cite{ColdP1,ColdP2}, and the $\nu=5/2$ fractional quantum Hall
state\cite{Nayak, ReadGreen}. Unfortunately, whether these systems
can be used to realize the elusive MFs is still a controversial
issue due to the lacking of convincing evidences. The conceptional
breakthrough in recent years is based on the idea that these exotic
quantum phases can be realized using the conventional $s$-wave
interaction, spin-orbit coupling (SOC) and Zeeman field\cite{Fu,
DSarma1, DSarma2, Alicea}. Very recently some pioneering experiments
both in one-dimensional (1D) nanowires and iron chain on the realization
of MFs has been carried out\cite{Exp1,Exp2, Exp3, Exp4, Exp5}. Remarkably, the localized wave functions at
the two ends with zero eigenenergy have been identified\cite{Exp5},
which provide important evidence for the realization of MFs. These
systems belong to topological D
 class from the Cartan classification due to absence of time-reversal (TR) symmetry\cite{Schnyder08,Kitaev}.
 In all these systems {\it only} the topological protected gapped phase can be realized, while the topological phase
transition is characterized either by the Chern number of the occupied bands, or somewhat equivalently
 the Pfaffian (Pf) of the Hamiltonian at the particle-hole (PH) invariant point(s)\cite{Kitaev1dd}.

\begin{figure}
        \centering
       \includegraphics[width=2.5in]{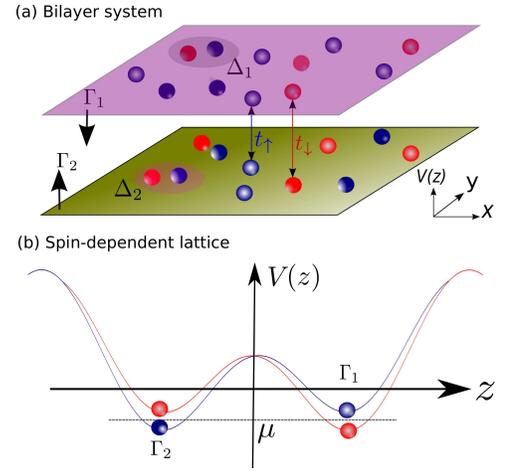}
       \caption{{\bf Bilayer system for TR invariant superfluids}. (b) The bilayer structure
can be realized by imposing a strong double-well potential along the
$z$ direction. The possibility of individual tunability of the local
Zeeman fields, spin-dependent tunnelings and the SOC coefficients
can be used to realize different topological models with TR
symmetry. (b) Realization of local Zeeman fields with opposite sign
in the bilayer system for $T_{xy}$ model via a spin-dependent lattice;
see more details in Methods.} \label{fig-fig1}
\end{figure}

These progress strongly stimulates the persuit of topological superconducting phases
with TR symmetry. However, these fantastical phases generally involve some {\it unconventional}
$p_x+ip_y$\cite{Qi09}, $d_{x^2-y^2}$\cite{FuL,Sato10, Nakosai12, Nakosai, KTLaw} and
$s_{\pm}$\cite{shusa,zhangfan2,zhangfan1} pairings, which are
implausible in realistic materials. Therefore we have the 
intriguing question that whether there phases can exist with only 
$s$-wave interaction? Here, instead of seeking for other new materials, we explore
these phases in a synthetic bilayer system
 where the bilayer freedom enables the definition of nonlocal TR symmetry
between the layers. While the single layer physics is well-known in
some sense,  we show that the interlayer tunnelings and TR symmetry
can lead to fundamentally different superconducting phases. We
explore the topological protected gapped and gapless phases in both
DIII and BDI classes in this platform, in which the fully gapped phase
in DIII phase is characterized by $\mathbb{Z}_2$, while all the gapless phases
are characterized by nontrivial winding numbers \cite{yxzhao, beri}
around the topological defects --- the Dirac cones --- in
momentum space and are also manifested from the Majorana flat bands
(MFBs)\cite{satofb}. We have proposed a scheme to realize
some of these phases in experiments. These results are confirmed using
detailed self-consistent calculation, in which the order parameters
and chemical potential are determined by the extrema of the free
energy. We also unveil the intimate relation between TR and mirror
symmetries in our models due to phase locking effect between the
pairings in the two layers, from which we realize some mirror
symmetry protected gapped and gapless phases. We finally demonstrate that these 
phases will not be spoiled by the interlayer pairings. We expect these new 
phases to greatly enrich our understanding of topological superconducting phases and
related transitions.

{\bf Results}

{\bf Physical Model}. We first briefly discuss the topological
superfluids in a single layer system, which has been intensively
studied in recent years\cite{MGong11, MGong12, HH11, ZQY,
He12}. In this model the Zeeman field is used to break the Kramers
degeneracy at zero momentum, so that the chemical potential can fill
just one band, around which an effective $p$-wave pairing can be
induced. We found that the topological phase can be realized in a
narrow parameter regime even at finite temperature though the Zeeman
field is (much) larger than the pairing strength\cite{MGong11,
MGong12}, so the induced $p$-wave pairing provides a new mechanism
to go beyond the Chandrasekhar-Clogston limit\cite{CC1,CC2}. The
recent interests in this system also include the topological inhomogeneous
superfluids\cite{FF1, FF2, huhuiff}, the quench dynamics of the
topological superfluids\cite{DongY, Victor} and the topological
superradiant state\cite{Jian}. Thus the exploring of these exotic
phases in this model is still an important topic in modern research.

To explore an experimental realization of the topological phases with TR
symmetry, we consider a two-dimensional (2D) bilayer system, see the schematic
structure in Fig. \ref{fig-fig1}, which can be realized by imposing
a strong double-well potential along the $z$ direction\cite{DW1, DW2}.
This synthetic structure has been widely used in recent literatures
for realizing some exotic phases in solid systems\cite{BL1, BL2}. We
first consider the case that both layers have identical SOC coefficients,
$V_{\text{so}}^{i} = ( s_i k_y\sigma_x-\lambda_i k_x \sigma_y )$,
where $\lambda_i = \alpha'$ and $s_i = \alpha$, $i=1, 2$, are the
SOC strength and $\sigma_\alpha$ are the Pauli matrices, and
identical spin-dependent tunnelings $t_{\uparrow,\downarrow} =t$,
but with different Zeeman fields $\Gamma_i$. The SOC in ultracold
atoms both for fermions and bosons has been extensively explored
using two coupled Raman
beams\cite{Lin11,Zhang12,Wang12,Cheuk12,Qu13,Chris14}, and very
recently, the required 2D SOC has been realized in
$^{40}$K\cite{Zhang15}. The barrier width between the two layers can
be controlled by the laser intensity in experiments thus we can
safely assume that only the intralayer pairings, which can be
defined as $\Delta_i = \sum_{\bf k} g\langle c_{i{\bf k}\uparrow}
c_{i-{\bf k}\downarrow}\rangle$, are important. Here $g$ is the
effective 2D interaction strength, $c_{i{\bf k}s}$ is the annihilate
operator for particle with momentum $\textbf{k}$ and spin $s$. The
corresponding Bogoliubov-de Gennes (BdG) equation can be written as
\begin{equation}
        \mathcal{H} = \begin{pmatrix}
                \mathcal{H}_0({\bf k})  & \bar{\Delta} \\
                                        \bar{\Delta}^\dagger   &
                                        -\mathcal{H}_0^*(-{\bf k})
        \end{pmatrix},
\label{eq-H}
\end{equation}
with $\mathcal{H}_0({\bf k})$ being the single particle Hamiltonian
\begin{equation}
        \mathcal{H}_0({\bf k}) = \begin{pmatrix}
                \epsilon_{{\bf k}} + \Gamma_1  & \rho_{1{\bf k}}  & -t_{\uparrow}  & 0  \\
                \rho_{1{\bf k}}^*  & \epsilon_{{\bf k}} - \Gamma_1 & 0  & -t_{\downarrow}  \\
                -t_{\uparrow}  &  0  & \epsilon_{{\bf k}} + \Gamma_2  & \rho_{2{\bf k}}   \\
                           0   & -t_{\downarrow} & \rho_{2{\bf k}}^*  & \epsilon_{{\bf k}} - \Gamma_2
                \end{pmatrix}.
\label{eq-H0}
\end{equation}
Here $\epsilon_{{\bf k}} = {\bf k}^2/2m - \mu$, and $\rho_{i{\bf k}}
= s_i k_y + i\lambda_i k_x$, and the concrete form of $\bar{\Delta}$
can be found in Methods (Eq. \ref{eq-Delta}). We construct the BdG
Hamiltonian in the Nambu spinor, $\Psi = (c_{1{\bf k}\uparrow},
c_{1{\bf k}\downarrow},c_{2{\bf k}\uparrow}, c_{2{\bf k}\downarrow},
c_{1{\bf -k}\uparrow}^\dagger, c_{1{\bf
-k}\downarrow}^\dagger,c_{2{\bf -k}\uparrow}^\dagger, c_{2{\bf
-k}\downarrow}^\dagger)^T$. Notice that in ultracold atoms the order
parameters and chemical potential should be solved
self-consistently.

We define the TR operator $\mathcal{T}$ for the single particle
Hamiltonian $\mathcal{H}_0$ in Eq. \ref{eq-H0} and $\mathbb{T}$ for
the BdG equation in Eq. \ref{eq-H}, while both $\mathcal{T}^2 = -1$
and $\mathbb{T}^2=-1$ are required for the topological superfluids
in DIII class. The realization of topological BDI class superfluids
with $\mathcal{T}^2 = +1$ and $\mathbb{T}^2=+1$ will be discussed
later. Obviously, without the Zeeman fields, this model has a
nonlocal TR operator $\mathcal{T} = \rho_x \otimes \sigma_yK$, where
$\rho_\alpha$ are Pauli matrices acting on the layer space and $K$
is the complex conjugate operator. In this TR operator, the spin up
(down) in one layer with momentum ${\bf k}$ is mapped to spin down
(up) in the other layer with opposite momentum. This symmetry is
still respected by the Zeeman fields when $\Gamma_1 = -\Gamma_2 =
\Gamma$, which is the basic reason for the local Zeeman fields
discussed before. These local Zeeman fields may be challenging to be
realized in solid state system, however, in ultracold atoms --- in the ideal
condition --- it can be realized using the following spin-dependent optical lattice,
\begin{equation}
V(z) = V_{\text{e}}(z) + V_{\text{o}}(z) \sigma_z,
\label{eq-Vz}
\end{equation}
where the subscripts e and o represent an even and odd function of
$z$, respectively. The first term describes the spin-independent
symmetric double-well potential, which has been realized in
experiments\cite{DW1, DW2}, and the second term means that the two layers
have {\it exactly} opposite Zeeman fields. Such a structure can be realized
using four counterpropagating laser beams (see details in Methods). An optical lattice that
slightly deviates from the above scenario can still be used to realize the model
with TR symmetry by fine-tuning the parameters.

The TR symmetry for the BdG equation is defined as $\mathbb{T} =
\text{diag}(\mathcal{T}, e^{i\phi}\mathcal{T})$, where $\phi$ is an
arbitrary phase. The TR symmetry requires that $\mathcal{T}
\bar{\Delta} \mathcal{T}^{-1}e^{-i\phi} =\bar{\Delta}$, namely,
$\Delta_1 = \Delta_2^*e^{-i\phi}$. The phase $\phi$
is used to compensate the global phase of the order parameters such that
only the relative phase between the two order parameters
$\Delta_{1,2}$ is important. After self-consistent calculation (see
details in Methods) we find that the two order parameters can be
treated as real numbers simultaneously. More precisely, we find that the
two layers can have zero relative phase and $\Delta_1/\Delta_2 > 0$
when the two tunnelings have the same sign; otherwise, the relative
phase is $\pi/2$ and $\Delta_1/\Delta_2 < 0$. This effect is rather
robust, and this feature is independent of other parameters such as binding
energy, Zeeman fields {\it etc}. We further find that the relative
phase between the two order parameters can only be introduced to the
Hamiltonian by the complex tunneling terms. Hereafter, this effect is called 
the phase locking effect. For these reasons, throughout this work, we treat the two order 
parameters as real numbers in the numerical simulation. For this particular model we have
\begin{equation}
\mathbb{T} = \Lambda K = \tau_0 \otimes \mathcal{T},
\label{eq-T}
\end{equation}
where $\tau_0$ is a Paul matrix acting on PH space. The intrinsic PH operator is defined as $\mathbb{C} = \tau_x K$, which ensures that $\mathbb{C}
\mathcal{H}({\bf k}) \mathbb{C}^{-1} =- \mathcal{H}(-{\bf k})$.

Hereafter, this model is dubbed as $T_{xy}$ model to distinct itself
from the other models studied in the following.

The $\Lambda$ operator in Eq. \ref{eq-T} is essentially the mirror symmetry operator since in our model
only the $k_x$ component contains the imaginary number, which can change sign upon complex conjugation $K$.
Thus we can define the mirror symmetry operator as $\mathcal{M}_y = \Lambda$, where
\begin{equation}
\mathcal{M}_y \mathcal{H}(k_x, k_y) \mathcal{M}_y^{-1} = \mathcal{H}(k_x, -k_y).
\label{eq-My}
\end{equation}
We can also define another mirror operator about $k_y$-axis as $\mathcal{M}_x  = \mathbb{I} \cdot \mathcal{M}_y$,
where $\mathbb{I} = \tau_z \otimes \rho_0 \otimes \sigma_z$ is the
inversion symmetry operator, $\mathbb{I} \mathcal{H}(\textbf{k})
\mathbb{I}^{-1} = \mathcal{H}(-\textbf{k})$. In this case,
\begin{equation}
\mathcal{M}_x \mathcal{H}(k_x, k_y) \mathcal{M}_x^{-1}
=\mathcal{H}(-k_x, k_y).
\label{eq-Mx}
\end{equation}
Notice that both $\mathcal{M}_x$ and $\mathcal{M}_y$ are unitary hermitian operators and
their eigenvalues are either $-1$ or $+1$
since $\mathcal{M}_{x,y}^2 = 1$.

These new symmetries can fundamentally change the topological
invariant of the Hamiltonian in the mirror symmetric invariant 1D
subspaces along $k_x$ and $k_y$ axes. To this end, we can diagonalize the
mirror symmetry operators via a unitary transformation $
U_{\alpha}\mathcal{M}_{\alpha} U_{\alpha}^{-1} = (\bf{I}_4,
-\bf{I}_4)$, where $\bf{I}_4$ is a $4\times 4$ unity matrix
and $\alpha = x, y$. Under this transformation we have
\begin{equation}
U_{\alpha} \mathcal{H}_{\alpha} U_{\alpha}^{-1} =\text{diag}(M_1^{\alpha}, M_2^{\alpha}),
\label{eq-M12}
\end{equation}
where $\mathcal{H}_x = \mathcal{H}(0, k_y)$ and $\mathcal{H}_y =
\mathcal{H}(k_x, 0)$. In general $M_{1,2}^{x,y}$ can have totally
different symmetries than the original Hamiltonian $\mathcal{H}$.
The concrete form of these matrices and their topological invariants
can be found in Methods. For the $T_{xy}$ model considered here we
find that both $M_{1,2}^{x,y}$ belong to AIII class, while these two
matrices are connected by TR symmetry and PH symmetry to recover the
symmetries of the original Hamiltonian. In fact the consequence of
mirror symmetry for any general Hamiltonian has been intensively
studied in literatures\cite{MS1, MS2, Ueno, Ken} by defining
$\mathcal{M}_{ss'}^{\alpha}:= \mathcal{M}_{\alpha} \mathbb{T}
\mathcal{M}_{\alpha}^{-1} = s \mathbb{T},
\mathcal{M}_{\alpha}\mathbb{C}\mathcal{M}_{\alpha}^{-1} = s'\Sigma$,
where $s, s' = \pm 1$. This quantity, $\mathcal{M}_{ss'}^{\alpha}$,
is crucial to determine the possible symmetries of the mirror
symmetric invariant subspaces, which are fully consistent with our
direct calculations above. This method will be adopted to find {\it
all} the possible mirror symmetric invariant Hamiltonians and their
topological classes for all the models presented in Table
\ref{tableI}.

\begin{figure}
           \centering
           \includegraphics[width=2.8in]{Txy.eps}
           \includegraphics[width=2.8in]{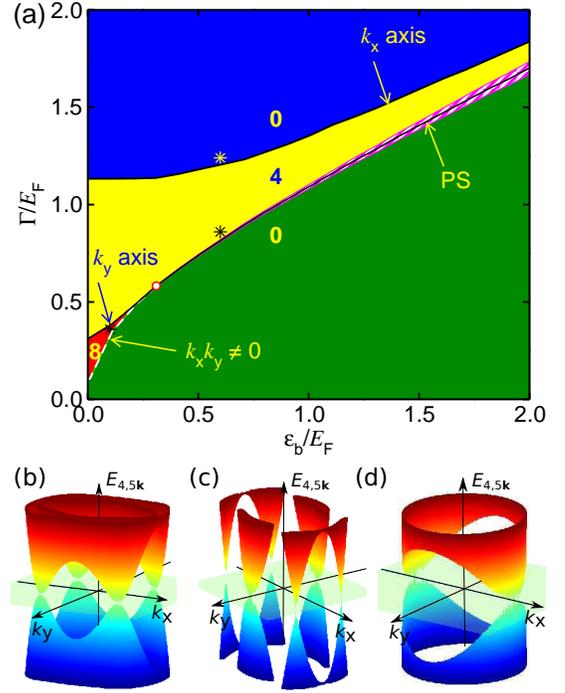}
\caption{\textbf{Topological superfluids for
$T_{xy}$ model}. (a) The phase diagram in the parameter space by $\epsilon_{\text{b}}$ and $\Gamma$,
in which the order parameters and chemical potential are determined self-consistently. The number in
each colored regime marks the number of gapless Dirac cones in momentum
space, thus $0$ defines a fully gapped phase. The two solid line
means that the energy gap closing and reopening take place at $k_x$
and $k_y$ axes. The dashed line means gap closing and reopening
at nonzero nonzero momenta ($k_xk_y\ne 0$). PS denotes the regime
for phase separation, which is visible when $\epsilon_{\text{b}} >
0.3E_{\text{F}}$. Parameters are: $t=0.6E_{\text{F}}$,
$\alpha=1.0E_{\text{F}}$, $\alpha'=0.5E_{\text{F}}$. (b) - (d) are
band structures at the critical boundaries when $\epsilon_\text{b} =
0.1E_\text{F}$.} \label{fig-fig2}
\end{figure}

{\bf Topological phases in $T_{xy}$ model}. We plot the phase
diagram as a function of binding energy $\epsilon_\text{b}$ and
Zeeman field $\Gamma$ in Fig. \ref{fig-fig2}a, in which the number
in each colored regime represents the number of robust Dirac cones.
We find that the two fully gapped phases are separated by two
gapless phases. The origin of these gapless phases can be understood
using the following way. Due to the presence of both PH and TR
symmetries, we can define a unitary sublattice symmetry $\mathbb{S}
= \mathbb{T} \cdot \mathbb{C}$, which satisfies $\{\mathbb{S},
\mathcal{H}\} =0$. We then diagonalize this sublattice operator via
a unitary transformation $V\mathbb{S}V^{-1} = \text{diag}(\bf{I}_4,
-\bf{I}_4)$, under which we have
\begin{equation}
V\mathcal{H}V^\dagger = \begin{pmatrix}
0 & Q \\
Q^\dagger & 0
\end{pmatrix}, \quad \text{Det} \mathcal{H} = |\text{Det}(Q)|^2.
\label{eq-Qxy}
\end{equation}
The closing of energy gap means that both the real and imaginary part of Det$(Q)$ should
be zero simultaneously, which give the following two conditions,
\begin{eqnarray}
(\boldsymbol{\gamma}\cdot {\bf k})^2 =t^2, \quad \epsilon_{{\bf k}}^2 = \Gamma^2 - \Delta^2,
\label{eq-Txy}
\end{eqnarray}
where $\boldsymbol{\gamma} = (\alpha', \alpha)$ and $\Delta_1 =
\Delta_2=\Delta$. The first equation defines an ellipse and the
second equation defines one circle or two circles, depending
strongly on the values of the parameters. The robust gapless phases
thus are determined by the overlaps between the ellipse and the
circle(s), which may always have intersections in some appropriate
parameter regimes. A topological explanation for the robust
accidental degeneracy at the Dirac cones and the physical meaning of
Eq. \ref{eq-Txy} will be presented shortly later. When $t \ne 0$, we
find the first equation prohibits the gap closing and reopening at
zero momentum, thus the topological phase transitions in this model
will always take place at nonzero momentum (${\bf k} \ne 0$). This
is in sharp contrast to the results in topological D class
superconducting phases\cite{Exp1, Exp2, Exp3, Exp4, Exp5, MGong11, MGong12}
where the critical boundary is only determined by the gap closing and reopening
at ${\bf k}=0$. When $t=0$, the critical boundary is reduced to,
\begin{equation}
{\bf k}=0, \quad \Gamma^2=\mu^2+\Delta^2,
\label{eq-Gammac}
\end{equation}
which is well-known in previous literatures\cite{Exp1, Exp2, Exp3, Exp4, Exp5, MGong11, MGong12}.

\begin{figure}
\centering
\includegraphics[width=2.8in]{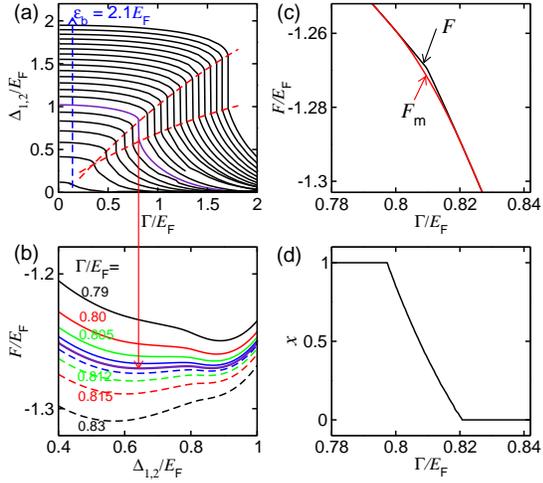}
\caption{{\bf Phase separation (PS) in $T_{xy}$ model}.
(a) The evolution of order parameters as a function of Zeeman field and binding
energy. The regime for sudden jump of order parameters is denoted by the two
dashed lines, which happens when $\epsilon_{\text{b}} > 0.3E_{\text{F}}$. (b)
The origin of the sudden jump of order parameters from the perspective of free energy $F$ using
$\epsilon_{\text{b}} = 0.8E_{\text{F}}$. (c) The mixed free energy $F_\text{m}$ (defined in
Eq. \ref{eq-mf}) can have slightly lower energy than $F$ by mixing of two different
superfluids. (d) The mixing coefficient $x$ as a function of Zeeman field. For parameters
see Fig. \ref{fig-fig2}.}
\label{fig-fig3}
\end{figure}

The phase diagram in Fig. \ref{fig-fig2}a can then be understood
from how the ellipse intersects with the circle(s). Starting from
the fully gapped trivial phase at $\Gamma \sim 0$ and gradually
increases the Zeeman fields, we find that the ellipse will first
intersect with two circles in the small binding energy regime
($\epsilon_{\text{b}} < 0.3E_{\text{F}}$), in which we observe a
topological gapless phase with 8 Dirac cones. This topological phase
transition will not take place at the $k_x$ and $k_y$ axes (see Fig.
\ref{fig-fig2}b). With the further increasing of Zeeman fields, one
of the circles will disappear when $|\mu| <
\sqrt{\Gamma^2-\Delta^2}$ and the system transits to the topological
gapless phase with 4 Dirac cones when $\epsilon_{{\bf
k}=(t/\alpha,0)}^2\epsilon_{{\bf k} = (0, t/\alpha')}^2 < (\Gamma^2
- \Delta^2)^2$ with gap closing at the $k_y$ axis for the particular
parameters used here (see Fig. \ref{fig-fig2}c). Finally, this phase
will evolve to the fully gapped phase with gap closing and reopening
along the $k_x$ axis (see Fig. \ref{fig-fig2}d). In the large
binding energy regime, however, only one circle is allowed in the
second equation of Eq. \ref{eq-Txy} so the fully gapped trivial
phase directly enters the phase with 4 Dirac cones. However, this
transition will be accompanied by a sudden jump of the order
parameter during the topological phase transition, see Fig.
\ref{fig-fig3}a. The corresponding jump of the chemical potential is
not shown. To understand this effect we calculate the free energy $F$ as a function of order
parameters in Fig. \ref{fig-fig3}b, in which two degenerate local
minima in $F$ are clearly shown in the critical boundary. 
The interplay of these two minima gives rise to the jump of parameters in Fig. \ref{fig-fig3}a.
The jump of these parameters is an important signal for phase separation (PS)\cite{caldas, WYI11}. 
Fortunately using the method in Refs. [\onlinecite{caldas, WYI11}] (see Eq.
\ref{eq-mf} in Methods) we find that the PS phase can only be
survived in a narrow regime near the boundary between trivial gapped phase and
the gapless phase with 4 Dirac cones.

\begin{figure}
\centering
\includegraphics[width=3in]{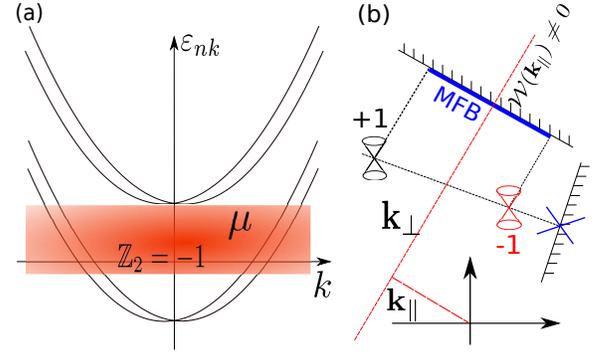}
\caption{{\bf Pictures for topological gapped superfluids and MFB.} (a) Topological gapped
superfluids with $\mathbb{Z}_2=-1$ can be realized when the chemical potential $\mu$
fills the lowest two bands, thus only one TR invariant point is enclosed by the Fermi
surface. This picture holds in the weak pairing limit. (b) The MFB is used to connect
the two defects with opposite chiralities ($\pm 1$),
which can be computed using $\mathcal{W}({\bf k}_{\parallel})$ in Eq. \ref{eq-Wp}. The
MFB is shrunk to a single point when the line connecting the two defects
perpendiculars to the strip direction.}
\label{fig-fig4}
\end{figure}

Next we try to characterize the topology of each phase. We first
focus on the fully gapped topological phase at strong Zeeman field,
which is characterized by $\mathbb{Z}_2$ from the Cartan
classification\cite{Schnyder08,Kitaev}. We employ the method
developed by Qi {\it et al}\cite{Qi10}, $\mathcal{N}_\text{2d} =
\prod_j \text{sgn}(\delta_{j{\bf k}})^{p_j}$, where $\delta_{j{\bf
k}} = \langle j \text{k} | i\mathcal{U} \Delta | j\text{k} \rangle $
is the pairing gap at the $j$th Fermi surface with
$|j\textbf{k}\rangle$ being the eigenfunction of $\mathcal{H}_0$,
and $p_j=1$ if the TR invariant point at ${\bf k}=0$ is enclosed by
the Fermi surface and $0$ otherwise. In this formula, $\mathcal{U} =
\rho_x\otimes \sigma_y$ and $i\mathcal{U}\bar{\Delta} = \rho_x\otimes {\bf I}_2$
is a Hermitian operator thus $\delta_{j{\bf k}} \in \mathbb{R}$. Notice that for a
relative large pairing strength in this regime, we can always
adiabatically deform this state to the weak pairing state near zero
binding energy or large Zeeman field without closing the energy gap,
thus this criteria can always be applied in this work to
characterize the topology of this fully gapped phase. A lengthy but
straightforward calculation shows that $\delta_{j{\bf k}} =
4\Delta (|t| \pm |\boldsymbol{\gamma}\cdot {\bf k}|)/(\Gamma \pm
\Upsilon_s)$, where $\pm$ defines the upper and lower two bands and
$\Upsilon_s = \sqrt{\Gamma^2 + (|t| + s|\boldsymbol{\gamma}\cdot {\bf
k}|)^2} \ne 0$. We immediately see from this result that
$\delta_{j{\bf k}}$ can change sign only when the energy gap is
closed at $t \pm |\boldsymbol{\gamma}\cdot {\bf k}|=0$, which is exactly the 
first equation in Eq. \ref{eq-Txy}. The second equation therefore determines the position of 
Fermi surface when $\Delta = 0$. Notice that the sign of $\delta_{j{\bf k}}$ 
is fixed in each Fermi surface. Then we find that in this regime, $\mathcal{N}_\text{2d} = -1$. In this regime
the chemical potential just fills the lower two bands while the upper two bands
are unoccupied (see Fig. \ref{fig-fig4}a). The same method yields $\mathcal{N}_\text{2d}
= +1$ in the fully gapped trivial phase with small Zeeman field.

\begin{figure}
\centering
\includegraphics[width=2.8in]{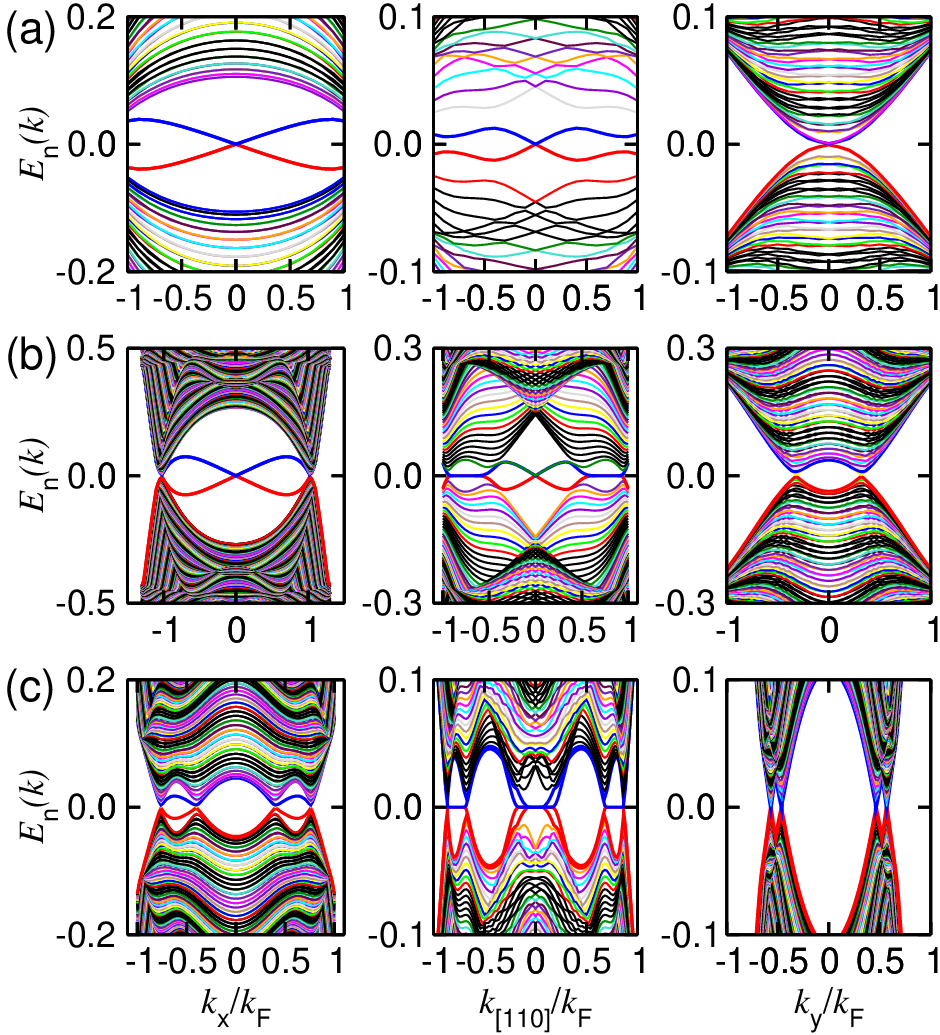}
\caption{\textbf{Edge states in $T_{xy}$ model along different directions}.
Edge states for a strip with $L=200/k_{\text{F}}$ along $k_x$ (left), $[110]$ (middle) and $k_y$
(right) directions in the fully gapped topological phase (a),
gapless phase with 4 (b) and 8 (c) Dirac cones for the three points marked by asterisk (*)
in Fig. \ref{fig-fig2}.  For parameters see Fig. \ref{fig-fig2}. }
\label{fig-fig5}
\end{figure}

This topological gapped phase can support edge states for
a strip along different directions; see Fig. \ref{fig-fig5}a. Note
that this gapped phase along $k_x$ and $k_y$ axes may also be
protected by mirror symmetry. For this reason, we calculate the
topological invariant --- the winding number in 1D --- in the mirror
symmetric invariant subspaces $M_{1,2}^{\alpha}$, which belong to
AIII class with only sublattice symmetry. This mirror winding number
$\mathcal{W}^{\alpha}$ is calculated using Eq. \ref{eq-Wp} (see Methods), which are
$\mathcal{W}^{x}=1$ and $\mathcal{W}^{y}=-1$. However, this mirror symmetry can
not protect the edge states along other directions.

The topological gapless phases emerged in our model can not be
characterized by the previous $\mathbb{Z}_2$ index. Instead, we
should use the winding numbers, or equivalently the chiralities, to characterize these
topological defects in momentum space. In the vicinity of the
defect, the effective Hamiltonian can be written as $H_{\text{eff}}
= \sum_{i,j=x,y} v_{ij}({\bf k})\delta k_i\sigma_j$, where ${\bf k}$
denotes the position of the defect. The chirality thus is defined as
$\nu({\bf k}) = \text{sgn} (\text{Det} [v]) = \pm 1$, where $[v]$ is
a matrix constructed by matrix elements $v_{ij}$. We may also
calculate the winding number $\mathcal{W}({\bf k})$ using Eq.
\ref{eq-Wo} for the BdG equation by considering a loop $l: |{\bf k-q}| = r$ around the
topological defect. We have a general relation between these two
invariants, $\mathcal{W}({\bf k}) = -\nu({\bf k})$ (see Eq. \ref{eq-Wnu} in Methods).
The TR symmetry plays a crucial role here
because the two topological defects with opposite momenta ($\pm {\bf
k}$) should have the same chiralities and winding numbers. These defects can only
be annihilated by fusing two defects with opposite chiralities.
The neutrality of topological charge in the entire space means that some
other two topological defects with opposite chiralities should be
presented at some other momenta. For this reason, the system can
only admit gapless phases with 4 or 8 Dirac cones. It is worth to
emphasize that these gapless phases may be a general feature in all
topological superconducting phases\cite{zhangfan1, zhangfan2, shusa}.

The edge states in these gapless phases will exhibit some salient anisotropic
features along different directions; see Fig. \ref{fig-fig5}b. When
the strip is along $[110]$ direction, we observe two MFBs at finite
momenta and one in-gap gapless linear excitation near zero momentum. The
two MFBs are a typical feature in gapless phases\cite{zigzag},
which are used to connect the two Dirac cones with opposite
chiralities. These MFBs are protected by topology thus is robust
against perturbations. To this end, we can calculate the winding
number $\mathcal{W}(\textbf{k}_{\parallel})$ along an infinity
straight line perpendicular the strip direction using Eq.
\ref{eq-Wp} (see Fig. \ref{fig-fig4}b), where $\textbf{k}_{\parallel}$ is the momentum along the strip
direction. We find that $\mathcal{W}(\textbf{k}_{\parallel}) = -1$, which
disappears only when a topological defect is encountered at Det$Q({\bf k})=0$.
When the strip is along $k_x$ and $k_y$ directions, the Dirac cones
with opposite chiralities are projected to the same points, thus the MFBs are shrunk
to some single gapless points.

The second salient feature is about the topological invariant for a
fully gapped 1D line in momentum space, which is also characterized
by $\mathbb{Z}_2$. The definition of this topological index is
identical to that in $\mathcal{N}_\text{2d}$\cite{Qi09}. So in the gapless
phase regime where the $\mathbb{Z}_2$ index in 2D is ill-defined, we
can still define the $\mathbb{Z}_2$ index along some particular
directions when the gapless defects are carefully avoided. We find
that $\mathcal{N}_{\text{1d}} = -1$ along both $k_x$ and $[110]$
directions, thus for a strip along these directions we can observe
some in-gap linear excitations near zero momentum as demonstrated in
Fig. \ref{fig-fig5}b. We also find $\mathcal{N}_{\text{1d}} = +1$
when along $k_y$ direction since the gap closing and reopening in
the 2D bulk takes place at the $k_x$ axis ($k_y = 0$). As a
consequence the in-gap linear excitation near zero momentum is
absent. This result is consistent with the mirror winding number
analysis which are $\mathcal{W}^x = +1$ and $\mathcal{W}^y = 0$.

The same analysis can be applied to the regime with 8 Dirac
cones, where the major observations are quite similar to the other
gapless phase. However, in this regime it is possible to realize
$\mathcal{W}({\bf k}_{\parallel}) = 2$ along some particular directions --- for instance
$[110]$ direction --- with doubly degenerate MFBs near zero
momentum; see Fig. \ref{fig-fig5}c. Moreover, two  MFBs can be found
at nonzero momenta. We also found the $\mathbb{Z}_2$ invariant is
trivial ($\mathcal{N}_{\text{1d}} = +1$) for a strip along any
direction, while in the mirror invariant subspaces we find that the mirror
winding number is $\mathcal{W}^x = \mathcal{W}^y = 0$, so we can not
observe any gapless linear excitations near zero momentum in this
phase.

While some of these phases are protected by both TR and mirror
symmetries, we need to point out that the basic role of mirror
symmetry played in this model is quite subtle. In the topological
gapped phase we indeed find that the winding number in the
mirror invariant subspaces are nonzero. Nevertheless if we
artificially break the mirror symmetry by introducing a relative
phase to the order parameters, we can still observe these edge
states. This means that the TR symmetry is a more important
protection in our model. However, in the BDI class phases below, we will
show that the mirror symmetry will play the primary
role in protecting the edge states in some parameter regimes.

\begin{figure}
           \centering
           \includegraphics[width=2.8in]{Tx0.eps}
           \includegraphics[width=2.8in]{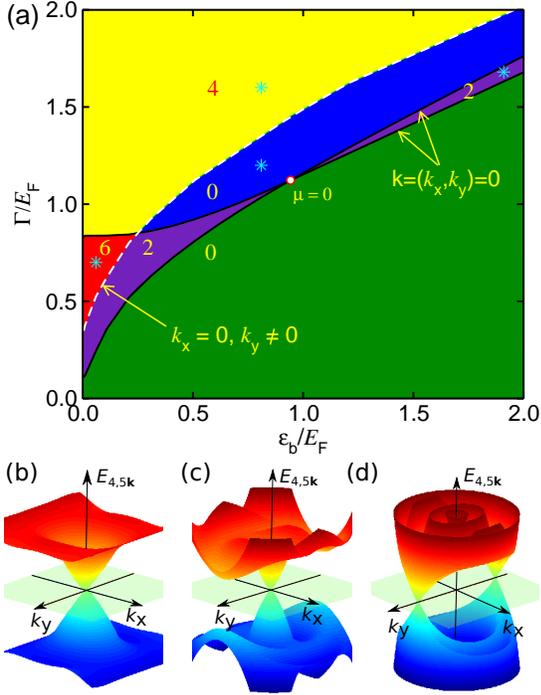}
\caption{\textbf{Topological superfluids for
$T_{x0}$ model}. (a) The phase diagram in the parameter space by $\epsilon_{\text{b}}$ and $\Gamma$.
The meaning of number is the same as that in Fig. \ref{fig-fig2}a. The two solid lines means that
the energy gap closing and reopening take place at ${\bf k} = 0$, and the
dashed line means the gap closing and reopening at $k_x=0$ and $k_y \ne 0$.
Parameters are: $t_{\uparrow,\downarrow} = 0.5E_{\text{F}}$,
$\alpha=\alpha'=1.0E_{\text{F}}$. (b) - (d) show the band structures
at the critical boundaries when $\epsilon_\text{b} = 0.5E_\text{F}$. }
\label{fig-fig6}
\end{figure}

\begin{figure}
\centering
\includegraphics[width=2.8in]{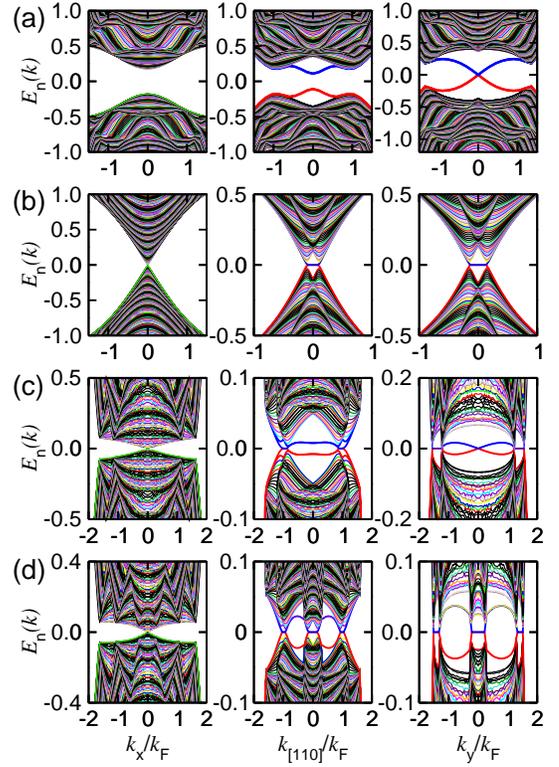}
\caption{\textbf{Edge states in $T_{x0}$ model along different
directions}. Edge states for a strip with width $L=200/k_{\text{F}}$ along $k_x$ (left), $[110]$
(middle) and $k_y$ (right) directions in the fully gapped topological
phase (a), gapless phase with 2 (b), 4 (c) and 6 (d) Dirac cones
for the four points marked by asterisk (*) in Fig. \ref{fig-fig6}.
For parameters see Fig. \ref{fig-fig6}.} \label{fig-fig7}
\end{figure}

{\bf Topological superfluids with other TR symmetries}. Now we try to extend this idea
to more general Hamiltonians by assuming that all coefficients --- $s_i$, $\lambda_i$, $\Gamma_i$
and $t_{\sigma}$
--- can be independently tuned in each layer. In this condition we can perform an
exhaustive search strategy to find all possible models with TR symmetry, which belong
to either DIII class with $\mathbb{T}^2 = -1$ or BDI class with
$\mathbb{T}^2 = +1$. We first assume that the single particle
$\mathcal{H}_0$ has time-reversal symmetry $\mathcal{T} =
\mathcal{U} K$, where $\mathcal{U} = \rho_i \otimes \sigma_j$ and
$\mathcal{T} \mathcal{H}_0({\bf k}) \mathcal{T}^{-1} =
\mathcal{H}_0(-{\bf k})$. With this operator, we can construct the
TR operator for the BdG Hamiltonian, which can be defined as either
$\mathbb{T} = \tau_0 \otimes \mathcal{U} K$ or
$\mathbb{T} = \tau_z \otimes \mathcal{U} K$. For convenience this
system is dubbed as $T_{ij}$ model. All the possible results using
this method are summarized in Table \ref{tableI}, in which only the
models with nonzero tunnelings and nonzero SOC coefficients are
presented. We find ten different models with TR symmetries, six of
which can support nontrivial topological phases. Notice that some
of these models require spin-dependent tunnelings, which may be
realized by spin selective laser-assisted hopping between the two
layers\cite{LAH} and by modulating the spin-dependent optical
lattice\cite{shaking1,shaking2}. The two layers with different SOC
coefficients may need some more complicated laser configurations.
There is an interesting relationship between the sign of SOC coefficients
and spin-dependent tunnelings, namely, a sign flip for the SOC
coefficients $s_1/s_2$ and $\lambda_1/\lambda_2$ is equivalent to a
sign flip in $t_{\uparrow}/t_{\downarrow}$, which can be mapped to
each other by a unitary transformation. Therefore
the $T_{xy}$ and $T_{yx}$, $T_{xx}$ and $T_{yy}$, and $T_{x0}$ and
$T_{xz}$ are mathematically equivalent, although their physical
realizations are totally different. Here we are interested in these
models because all these phases may in principle be realized based
on $s$-wave interaction in this versatile platform, where different
phases are possible to be tuned to each other.

\begin{table*}[t]
\centering \caption{\textbf{Symmetry table for all the possible
superfluids with TR symmetry}. The single particle TR operator is
defined as $\mathcal{T} = \mathcal{U} K$, where $\mathcal{U}$ is
shown in the second column. The relation between the coefficients
$\Gamma_i$, $s_i$, $\lambda_i$, $t_i$ and $\Delta_i$ are shown from
column three to seven, in which the asterisk (*) means that their values can be
arbitrary real numbers while $0/0$ means that both the two
parameters should be zero simultaneously. $\mathbb{T}$ is the TR
operator for the BdG Hamiltonian and $\mathcal{M}_{ss'}^{\alpha}$
defines the symmetry in the mirror invariant subspace. The last
column shows whether topological gapped and/or gapless phases can
exist (Y) or not (N) in some appropriate parameter regimes.}
\begin{tabular}[t]{p{0.07\textwidth} p{0.09\textwidth} p{0.07\textwidth}
p{0.07\textwidth} p{0.07\textwidth} p{0.07\textwidth}
p{0.07\textwidth} p{0.12\textwidth}  p{0.12\textwidth}
p{0.12\textwidth} p{0.04\textwidth}} \hline Model  & $\mathcal{U}$
&$\Gamma_1/\Gamma_2$  & $s_1/s_2$  & $\lambda_1/\lambda_2$  &
$t_\uparrow/t_\downarrow$  & $\Delta_1/\Delta_2$  & $\mathbb{T}$
($\mathbb{T}^2$) & Mirror $\mathcal{M}_{ss'}^{x}$ & Mirror
$\mathcal{M}_{ss'}^{y}$ & Topo.  \\  \hline
$T_{xx}$ & $\rho_x \otimes \sigma_x$  & -1  & -1  & -1  & +1  & +1 &  $\tau_z\otimes \mathcal{T}$ (+1)   & $\mathcal{M}_{+-}^x$ (AI) & $\mathcal{M}_{+-}^y$ (AI) & N   \\
$T_{yy}$ & $\rho_y \otimes \sigma_y$  & -1  & +1  & +1  & -1  & -1 &
$\tau_z\otimes \mathcal{T}$ (+1)   & $\mathcal{M}_{+-}^x$ (AI) &
$\mathcal{M}_{+-}^y$ (AI) & N \\  \hline $T_{xy}$ & $\rho_x \otimes
\sigma_y$  & -1  & +1  & +1  & +1  & +1 &  $\tau_0\otimes
\mathcal{T}$ (-1)   & $\mathcal{M}_{--}^x$ (AIII)  &
$\mathcal{M}_{--}^y$ (AIII)& Y  \\
$T_{yx}$ & $\rho_y \otimes \sigma_x$  & -1  & -1  & -1  & -1  & -1 &
$\tau_0\otimes \mathcal{T}$ (-1)   & $\mathcal{M}_{--}^x$ (AIII)  &
$\mathcal{M}_{--}^y$ (AIII) & Y \\  \hline $T_{xz}$ & $\rho_x
\otimes \sigma_z$  & +1 & +1 & -1 &*  & -1 & $\tau_0\otimes
\mathcal{T}$ (+1)   & $\mathcal{M}_{+-}^x$ (AI)  &
$\mathcal{M}_{++}^y$ (BDI) & Y  \\
$T_{x0}$ &$\rho_x \otimes \sigma_0$ & +1 & -1 & +1 & *  & +1 &
$\tau_0\otimes \mathcal{T}$ (+1)   & $\mathcal{M}_{+-}^x$ (AI)  &
$\mathcal{M}_{++}^y$ (BDI)& Y   \\  \hline $T_{zy}$ &$\rho_z \otimes
\sigma_y$ &0/0 &  * & *  & -1 & *  & $\tau_0\otimes \mathcal{T}$
(-1)   & $\mathcal{M}_{--}^x$ (AIII)  & $\mathcal{M}_{--}^y$ (AIII) & N  \\
$T_{0y}$ &$\rho_0 \otimes \sigma_y$ &0/0 &  * & *  & +1 & *  &
$\tau_0\otimes \mathcal{T}$ (-1)   & $\mathcal{M}_{--}^x$ (AIII)  &
$\mathcal{M}_{--}^y$ (AIII) & N \\ \hline $T_{0z}$ &$\rho_0 \otimes
\sigma_z$ & *  &  * &0/0 & *  & *  & $\tau_z\otimes \mathcal{T}$
(+1)   &
--- & $\mathcal{M}_{+-}^y$ (AI) & Y  \\
$T_{00}$ &$\rho_0 \otimes \sigma_0$ & *  &0/0 &  * & *  & *  &
$\tau_0\otimes \mathcal{T}$ (+1)   & $\mathcal{M}_{+-}^x$ (AI)  &
--- & Y
 \\ \hline
\end{tabular}
\label{tableI}
\end{table*}

In all these models, our detailed self-consistent calculation demonstrates
that the intimate relation between TR and mirror symmetries is
always respected due to the phase locking effect. We determine all
the topological classes for the Hamiltonians in the mirror symmetric
invariant subspaces using the method in Eq. \ref{eq-M12}, which
are summarized in Table \ref{tableI}. The analysis based on
the parameter $\mathcal{M}_{ss'}^{\alpha}$ is completely consistent
with the direct analysis from the Hamiltonians\cite{MS1, MS2, Ueno, Ken}.
In all these models the subspaces can belong to AI, AIII and/or BDI
classes, where the later two classes can support nontrivial topological phases
in 1D\cite{Schnyder08,Kitaev}.

We utilize the $T_{x0}$ model as an example to highlight the unique features
of topological superfluids in BDI class. This is also a physical
model that may support topological protected MFs in 1D\cite{Sumanta}.
Unfortunately, this physical model may not have counterpart in
solid state systems with 2D SOC, so the realization of topological gapped and gapless
superfluids in BDI class is an important extension of our understanding of topological phases.
For this model the TR symmetry requires $s_1 =-s_2 =\alpha$, $ \lambda_{1,2} = \alpha'$ and
$\Gamma_{1,2} = \Gamma$ (see table I).
Hereafter we focus on the case $t_{\uparrow,\downarrow}=t$. The phase diagram from the
self-consistent calculation is presented in Fig. \ref{fig-fig6}a. To understand the origin of
the gapless phases, we calculate the determinant of the Hamiltonian using the
method in Eq. \ref{eq-Qxy}, which can be written as
\begin{equation}
\text{Det}(Q) = \mathcal{G}({\bf k}) +8 i \alpha' k_x \epsilon_{{\bf k}}t\Delta,
\label{eq-Tx0}
\end{equation}
where $\mathcal{G}({\bf k}) =(\epsilon_{\bf k}^2 - t^2 - k_y^2 \alpha^2 -
k_x^2 \alpha'^2 - \Gamma^2 + \Delta^2)^2 - 4 t^2 (k_x^2 \alpha'^2 +
\Gamma^2) +4 (t^2 + k_y^2 \alpha^2 + k_x^2 \alpha'^2) \Delta^2$,
with $\Delta_1 = \Delta_2 = \Delta$. A direct calculation shows that
the gapless points can only happen along the $k_y$ axis ($k_x = 0$). Furthermore
we can verify easily that when $\Gamma =0$, $\text{Det}(Q)$ is
always greater than zero when $t\ne 0$, so these topological superfluids are also
driven by the Zeeman field. In fact we need to emphasize that this
is a quite general feature for {\it all} the topological phases
summarized in Table \ref{tableI}. Mathematically, the polynominal equation
$\text{Det}(Q) =0 $ can always have real solutions in some parameter
regimes. In our numerical simulation, we find that the critical
boundaries are also determined by three lines as a function of
binding energy and Zeeman field.  The two solid lines give
topological transitions with gap closing and reopening at ${\bf
k}=0$ and $ (t\pm \mu)^2+\Delta^2=\Gamma^2$; see Fig.
\ref{fig-fig6}b-c. Obviously these two critical boundaries are
reduced to the well-known result in Eq. \ref{eq-Gammac} when $t=0$. The
dashed line is the gap closing and reopening at finite momentum ($k_x =0, k_y \ne 0$); see 
Fig. \ref{fig-fig6}d. These critical boundaries are totally different from the $T_{xy}$ 
model shown in Fig. \ref{fig-fig2}, which can have profound influence on the topological
phases with slightly TR symmetry breaking; see below.

We find that in this model the order parameters and chemical
potential are always a smooth function of Zeeman field and binding
energy, so the PS phase discussed in Fig. \ref{fig-fig3} is not
shown up. However, this model can exhibit some interesting features
that not seen in the previous $T_{xy}$ model. We notice that when
$\mathbb{T}^2 = +1$ the two Dirac cones with opposite momenta have
opposite chiralities (see an exact proof in Methods), so in
principle, all the topological protected gapless superfluids with
even number ($\le 8$) of Dirac cones are admitted. Our crucial
observations thus are in order. ({\bf 1}) We first focus on the
fully gapped phases which in principle can not support topological
phases in 2D from the standard table\cite{Schnyder08,Kitaev}.
However, mirror symmetry protected phases are still allowed, which
can be characterized by the winding number defined in Eq.
\ref{eq-Wp} along the mirror symmetric invariant subspaces. For the
fully gapped phase at small Zeeman field ($\Gamma \sim 0$), we find
that the mirror winding numbers along $k_y$ axis are
$\mathcal{W}_1^y \oplus \mathcal{W}_2^y = 0 \oplus 0$. Notice that
the mirror symmetric subspaces along $k_x$ direction belong to AI
class thus is always trivial in 1D\cite{Schnyder08,Kitaev}. For this
reason we do not observe gapless excitation for a strip along any
directions (not shown in Fig. \ref{fig-fig7}). However, in the fully
gapped phase at strong Zeeman field, we find that the mirror winding
numbers are $\mathcal{W}_1^y\oplus \mathcal{W}_2^y = -1 \oplus -1$ 
and $\mathcal{W}({\bf k}_{\parallel} = k_y) = +1$, 
so we can observe robust gapless excitations for a strip along $k_y$
direction in Fig. \ref{fig-fig7}a. ({\bf 2}) In the left phase with
2 Dirac cones, the mirror winding numbers are $\mathcal{W}_1^y
\oplus \mathcal{W}_2^y = -1\oplus 0$. thus we observe robust MFBs
along both $k_y$ and $[110]$ directions in Fig. \ref{fig-fig7}b. The
MFB is shrunk to a single point for a strip along $k_x$ direction
since all the Dirac cones are located at the $k_y$ axis. ({\bf 3})
In the regime with 4 Dirac cones, the mirror winding numbers are
$\mathcal{W}_1^y\oplus \mathcal{W}_2^y = -1 \oplus -1$, which are
exactly the same as the topological gapped phase at strong Zeeman
field due to gap closing and reopening at $k_y\ne 0$. This explains
why the similar in-gap excitations in Fig. \ref{fig-fig7}a can also
be observed in Fig. \ref{fig-fig7}c. Meanwhile we can observe two
MFBs at finite momenta due to the same reason discussed in $T_{xy}$
model. ({\bf 4}) In the regime with 6 Dirac cones, the mirror
winding numbers are $\mathcal{W}_1^y\oplus \mathcal{W}_2^y = -1
\oplus 0$, which are the same as the left gapless phase with 2 Dirac
cones in Fig. \ref{fig-fig7}b. In this phase regime we can observe
three MFBs for a strip along $k_y$ and $[110]$ directions in Fig.
\ref{fig-fig7}d, all of which will shrink to a single point for a
strip along $k_x$ direction. ({\bf 5}) These mirror symmetry
protected phases will be destroyed by mirror symmetry breaking
terms, such as the relative phase between the order parameters.

{\bf Discussion and conclusion}. Here we investigate the topological
phases in ultracold atom systems due to their flexibility of
tuning all the parameters in a wide range in experiments, which are
challenging in solid state system. In Methods we have shown a concrete scheme
to realize the bilayer structure with exactly opposite Zeeman
fields. However, it is still very interesting to ask the basic
question that what will happen under weak TR symmetry breaking? In
this condition both the DIII and BDI class phases will be collapsed to D
class phases\cite{MGong11, MGong12}, in which the topological phase
transitions can be characterized by either Chern number or Pfaffian
aforementioned and all gapless phases will be destroyed. In the $T_{xy}$ model we find that all the phases
under weak TR symmetry breaking have Chern number $\mathcal{C} = 0$
and Pfaffian $\mathcal{\chi} = \text{sgn Pf}(\mathcal{H}(0)\tau_x) = +1$. This is due to
the fact that all the energy gap closings and reopenings in this
model take place at nonzero momenta, so the Pfaffian at zero
momentum is unchanged. However, the BDI model can exhibit totally
different behaviors due to the possibility of gap closing and
reopening at zero momentum. So we find that the trivial gapped phase
will have $\mathcal{C} = 0$ and $\mathcal{\chi} = +1$, and the regime with  2 and 6 Dirac
cones will have $\mathcal{C} = 1$ and $\mathcal{\chi} = -1$, and the fully gapped topological
phase and the regime with 4 Dirac cones will have $\mathcal{C} = 2$
and $\mathcal{\chi} = +1$. In general, $\mathcal{\chi} = (-1)^{\mathcal{C}}$\cite{HBB}. 
If all the parameters in experiments can be changed adiabatically, it is
possible to observe the topological transitions among DIII, D and
BDI classes without gap closing\cite{Nagaosa}.

We finally emphasize that these phases will not be spoiled by the presence of
interlayer pairings. We have included all the interlayer
pairings in our model (see Methods) and minimize the total free energy $F$ with
respect to these parameters. The detailed self-consistent simulation shows that
the TR symmetry is always respected. This is due to the phase locking effect
between the two layers  with real tunnelings. These pairings only
slightly modify the real and imaginary part of Det$Q$ discussed in Eq. \ref{eq-Txy}
and Eq. \ref{eq-Tx0}, thus Det$Q = 0$ can always have real solutions in some parameter
regimes. Instead, these pairings just slightly modify the topological boundaries without
qualitatively influence all the conclusions.

To conclude, the bilayer structure defined in this work provides an
experimentally controllable platform to realize new topological
protected phases in ultracold atom systems with TR symmetry that
belonging to DIII and BDI classes. These phases can be realized using the conventional $s$-wave
interactions, thus is in stark contrast to previous physical
proposals based on unconventional pairings. In these models, the fully gapped
phase in DIII class is characterized by $\mathbb{Z}_2$, while all the gapless phases in these two 
classes are characterized by winding numbers around the topological defects. The phase locking 
effect between the two layers ensures the realization of mirror symmetry protected gapped
and gapless phases. These new phases are expected to greatly enrich
our understanding of topological superconducting phases and related transitions. This novel
idea can also be applied to explore the MFs in 1D and the Weyl
superfluids in 3D, and their crossover from
the Bose-Einstein condensation in the strong coupling limit to the
Bardeen-Cooper-Schrieffer in the weak coupling limit\cite{MGong11, MGong12,
HH11, ZQY, He12},  which will be published elsewhere.

{\bf Methods}

{\bf Generation of local Zeeman fields in $T_{xy}$ model}. The
spin-dependent double-well potential in Eq. \ref{eq-Vz} is essential for the first four 
models in Table \ref{tableI}, which can be realized using the state-of-the-art techniques. 
Here we follow the basic idea in Ref. [\onlinecite{Ivan}].  For the cold atom optically 
pumped from the $|J=1/2\rangle$ manifold to the $|J'=3/2\rangle$ manifold and in the
large detuning limit the rank 2 polarizability tensor takes the
following form\cite{Ivan},
\begin{equation}
\alpha_{ij}(J\rightarrow J') = \tilde{\alpha} ({2\over 3}\delta_{ij} - {i\over 3}\epsilon_{ijk}\sigma_k),
\label{eq-alpha}
\end{equation}
where $\tilde{\alpha}$ is a constant that inversely proportional to the detuning.
For transparency we first consider two counterpropagating laser beams with momenta $\pm k$ and frequency
$\omega$ along $z$ direction. Then the electric field can be written as
\begin{equation}
{\bf E} = (E_1 {\bf e}_1 e^{i(kz+\phi)} + E_2 {\bf e}_2 e^{-ikz})e^{i\omega t},
\end{equation}
where ${\bf e}_{1,2}$ represent the polarization direction for the
two laser beams with field strength $E_{1,2}$, and $\phi$ is the
relative phase between the two laser beams. Let ${\bf e}_1 =
\cos({\theta \over 2}){\bf e}_x + \sin({\theta \over 2}) {\bf e}_y$
and ${\bf e}_2 = \cos({\theta \over 2}){\bf e}_x - \sin({\theta
\over 2}) {\bf e}_y$, where $\theta$ is the relative polarization
angle between the two laser beams, then the spin-dependent optical
lattice can be written as\cite{Ivan}
\begin{equation}
U(z) = {\bf E}^*\cdot \alpha \cdot {\bf E}= \tilde{\alpha}[{2\over 3}|{\bf E}|^2 + {i\over 3}
({\bf E}^* \times {\bf E}) \cdot \boldsymbol{\sigma}],
\label{eq-Uz}
\end{equation}
where in the second term only the $\sigma_z$ component is nonzero since ${\bf E}$ and
${\bf E}^*$ lie in the $x$-$y$ plane. Plugging the expression of ${\bf E}$ into Eq. \ref{eq-Uz} we find
\begin{eqnarray}
U(z) = U [2\cos(\theta)\cos(2kz + \phi) + \sin(\theta) \sin(2kz + \phi) \sigma_z],
\end{eqnarray}
where $U=-{2\over 3} \tilde{\alpha} E_1E_2$. Thus these two laser beams create a spin-dependent
periodic optical lattice along $z$ direction.

To realize an isolated double-well structure in Fig.
\ref{fig-fig1}a, we can apply two extra laser beams along the same
direction with momenta $\pm 2k$\cite{DW1,DW2}, polarizability
$\tilde{\alpha}'$, electric field strength $E_{1,2}'$ (hence
$U'=-{2\over 3} \tilde{\alpha}' E_1'E_2'$), polarization angle
$\theta'$ and relative phase $\phi'$. These four colinear laser
beams can create a spin-dependent potential in the following form,
\begin{eqnarray}
U(z) && = 2U\cos(\theta)\cos(2kz) + 2U'\cos(\theta')\cos(4kz + \delta \phi) +  \nonumber \\
       && (U\sin(\theta) \sin(2kz) +  U'\sin(\theta') \sin(4kz + \delta
       \phi))\sigma_z,
\end{eqnarray}
where a shift of $z$ by $-\phi/2k$ has been made and $\delta \phi = \phi' - 2\phi$.
We plot the band structure in Fig. \ref{fig-fig1}b using the above equation
with $\delta \phi = 0$, which is exactly the ideal potential presented
in Eq. \ref{eq-Vz}. In this potential the first part ---an even function of $z$
---gives the symmetric double-well potential while the second part --- an odd
function of $z$ --- gives the spin-dependent Zeeman fields. The two layers
therefore have exactly opposite Zeeman fields and this feature is independent of the
choice of other parameters. We also need to emphasize that a bilayer system with TR symmetry
can still be achieved even deviates from this idea condition because five
independent parameters --- $U$, $U'$, $\theta$, $\theta'$, $\delta \phi$ --- can be
tuned to ensure just two constraints --- $t_{\uparrow} = t_{\downarrow}$ and
$\Gamma_1 = -\Gamma_2$ --- in $T_{xy}$ model, which can always be fulfilled
in a vast range of parameters. Furthermore the external Zeeman field along $z$
direction, $B_z\sigma_z$, can also contribute to the recovery of TR symmetry.

{\bf Self-consistent calculation}. The pairings in ultracold atoms is induced by the $s$-wave scattering
between the particles
thus should be determined in a self-consistent manner. In the major numerical simulation, we only consider
the intralayer pairings by assuming that the interlayer pairings are sufficiently suppressed by the strong barrier between the two
layers (see Fig. \ref{fig-fig1}).
The thermodynamic potential $\Omega$ is defined as $Z = e^{-\Omega/T} = \text{Tr} e^{-H/T}$.
When $T \rightarrow 0$ we have
\begin{equation}
\Omega = {1\over 2} \sum_{\eta < 0, {\bf k}} (E_{\eta{\bf k}} + 2\epsilon_{1 {\bf k}} +  2
\epsilon_{2{\bf k}})  + {|\Delta_1|^2 \over g} + {|\Delta_2|^2 \over g},
\label{eq-Omega}
\end{equation}
where $\eta < 0$ means summation over all the occupied bands with
$E_{\eta{\bf k}} \le 0$. The condensation energy should be regularized
using the prescription: $g = \sum_{{\bf k}} 1 / ({\bf k}^2/m +
\epsilon_\text{b})$, where $\epsilon_\text{b}$ defines the binding
energy which can be controlled by the Feshbach resonance. The
corresponding free energy used in Fig. \ref{fig-fig3}b is defined as
$F = \Omega + n\mu$. The order parameters and chemical potential are
determined by the extrema of $F$: $\partial F /\partial
\Delta_i = 0$, $\partial F /\partial \Delta_i^* = 0$ and  $\partial
F /\partial \mu =0$.

In our simulation, the two order parameters are assumed to $\Delta_1
e^{i\delta \theta}$, and $\Delta_2 e^{-i\delta \theta}$
respectively, where $\Delta_{1,2}$, $\delta \theta$ $\in \mathbb{R}$. 
The global phase has been gauged out, so $\delta \theta$
determines the relative phase between the order parameters in the
two layers. There three parameters together with the chemical
potential are used to minimize the total free energy $F$. In our
numerical simulation we have assumed $k_\text{F} = \sqrt{n\pi}$,
where $n$ is the fixed total particle density, and
$E_\text{F}=k_\text{F}^2/2m$. In all our simulation the energy and
momentum are rescaled by $E_\text{F}$ and $k_\text{F}$,
respectively. We find that in all our numerical simulation $\delta
\theta =0$, which means that the order parameters between the two
layers always have fixed phase. This basic feature, dubbed as phase
locking effect throughout this work, is still respected by the
presence of interlayer pairings (see below).

{\bf Phase separation}. The instability of the uniform phases
towards the PS phase can be understood from the mixing of two different
phases, which is captured by the following mixed free energy\cite{caldas,WYI11},
\begin{equation}
F_\text{m} = x F(\mu, \Delta_1, \Delta_2) + (1-x)F(\mu, \Delta_1',
\Delta_2'), \label{eq-mf}
\end{equation}
where $0 \le x \le 1$ defines the mixing coefficient (see numerical
result in Fig. \ref{fig-fig3}d). The two phases should have the
same chemical potential but different pairing strengthes
$\Delta_{1,2}$ and $\Delta_{1,2}'$. We minimize the total mixed free
energy $F_\text{m}$ with respect to all these parameters to map out
the PS phase in Fig. \ref{fig-fig2}a. We find that this phase
is not favorable in the $T_{x0}$ model.

{\bf Topological invariants in the gapless phases}. Now we investigate the effect of TR, PH
and mirror symmetries on the
winding numbers and chiralities for the gapless phases.

(1) We first notice that the off-diagonal matrix $Q$ in Eq. \ref{eq-Qxy} and Eq. \ref{eq-Tx0} have the
following
properties\cite{Schnyder08, Kitaev}
\begin{eqnarray}
       \text{Det} [Q_{\text{DIII}}^T({\bf k})] && =
       \text{Det}[-Q_{\text{DIII}}(-{\bf k})], \quad \text{  and } \\
        \text{Det}[Q_{\text{BDI}}^*({\bf k})] && = \text{Det}[Q_{\text{BDI}}(-{\bf k})].
\end{eqnarray}
In these models the winding number around a single topological defect is defined as
\begin{eqnarray}
\mathcal{W}({\bf k})
&& = {i\over 4\pi} \oint_l d{\bf q} \text{Tr} \mathbb{S} \mathcal{H}^{-1} \partial_{\bf q} \mathcal{H} \nonumber \\
&& = {1\over 2\pi}\Im \oint_l d{\bf q} \partial_{\bf q}\text{Det}Q({\bf q}) \in \mathbb{Z},
\label{eq-Wo}
\end{eqnarray}
where ${\bf k}$ defines the position of the defect and the loop $l : = |{\bf q-k}|=r$ should be small enough
to enclose only one defect. We have
\begin{eqnarray}
        \mathcal{W}_{\text{DIII}}({\bf k}) = \mathcal{W}_{\text{DIII}}(-{\bf
        k}),
        \mathcal{W}_{\text{BDI}}({\bf k}) = - \mathcal{W}_{\text{BDI}}(-{\bf k}).
\label{eq-WDIII}
\end{eqnarray}
We see that the TR symmetry plays a fundamental role in deriving the above results. In fact when TR is
broken the $Q$ matrix is no longer well-defined and the topological defects are then destroyed. We do not
observe the robust Dirac cones in topological D class phases in all our models
in Table \ref{tableI} by introducing some TR symmetry breaking terms to Eq. \ref{eq-H}.

(2) Next we study the chirality of the topological defect. We assume
the defect is located at ${\bf k}$ with degenerate zero energy
eigenvectors $\psi_{\pm}$, where $\psi_+ = \mathbb{S} \psi_-$ due to
chiral symmetry. Based on these functions we can define two chiral
basis $\phi_{\pm} = \psi_+ \pm \psi_-$, which satisfy $\mathbb{S}
\phi_{\pm} = \pm \phi_{\pm}$. We assume that $H({\bf k}+\delta
{\bf k}) = H({\bf k}) + M(\delta {\bf k})$ in the vicinity of the
defect, where $M(\delta {\bf k}) =\delta k_x M_x + \delta k_y M_y$.
The chiral symmetry ensures that $\mathbb{S} M \mathbb{S}^\dagger =
-M$. With these results we readily have $\langle \phi_{\pm} |M
| \phi_{\pm} \rangle =0$, and the off-diagonal term, $\langle \phi_+
| M |\phi_-\rangle = \langle \psi_+|M|\psi_+\rangle -\langle
\psi_-|M|\psi_-\rangle -\langle \psi_+|M|\psi_-\rangle+\langle
\psi_-|M|\psi_+\rangle$. Notice that $\langle
\psi_+|M|\psi_+\rangle= -\langle \psi_-|M|\psi_-\rangle$ and
$\langle \psi_+|M|\psi_-\rangle=-\langle \psi_-|M|\psi_+\rangle$, so
the last two terms should be exactly imaginary numbers. Therefore we
can assume $\langle \phi_+ | M |\phi_-\rangle = a_1 \delta k_x + a_2
\delta k_y + i (b_1 \delta k_x + b_2 \delta k_y)$, where $a_{1,2}$,
$b_{1,2}$ $\in \mathbb{R}$. The effective Hamiltonian can be written as
\begin{equation}
        H_{\text{eff}} = \begin{pmatrix}
                0   & c_1 \delta k_x + c_2 \delta k_y \\
                *  & 0
        \end{pmatrix} = \sum_{ij}v_{ij}\delta k_i \sigma_j,
        \label{eq-Heff}
\end{equation}
where $c_i = a_i+ib_i$. The effective Hamiltonian constructed in this way
can fully embody the importance of TR symmetry in this model. The chirality for Eq. \ref{eq-Heff} is
defined as
\begin{equation}
\nu = \text{sgn}(\text{Det} [v]) = \text{sgn}(b_1a_2-b_2a_1).
\end{equation}
We also find a general relation between these invariants,
\begin{equation}
\mathcal{W}({\bf k}) = -\nu({\bf k}).
\label{eq-Wnu}
\end{equation}

We turn to study the topological charge at $-{\bf k}$, which
also have two degenerate eigenvectors $\psi_{\pm}'$ with $\psi_+' = \mathbb{S} \psi_-'$.
The PH and TR symmetries ensure that $\psi_-' = \mathbb{T} \psi_- = \eta \mathbb{C} \psi_+$,
and $\psi_+' = \mathbb{C} \psi_- = \eta \mathbb{T} \psi_+$, where $\eta = \mathbb{T}^2$.
The perturbation term is assumed to be $M'$. Let $\phi_{\pm}' = \eta \mathbb{T} \psi_+ \pm \mathbb{T} \psi_-$, 
then 
\begin{eqnarray}
        \langle \phi_+'|M'|\phi_-'\rangle = \langle \eta \psi_+ - \psi_- | M| \eta \psi_+ + \psi_-\rangle.
\end{eqnarray}
For DIII model, $\eta = -1$, and $\langle \phi_+'|M'|\phi_-'\rangle = \langle \phi_+|M|\phi_-\rangle$, so the two defects
with opposite momenta have the same chirality. They will have opposite chiralities when $\eta = 1$ in BDI model since
$\langle \phi_+'|M'|\phi_-'\rangle = (\langle \phi_+|M|\phi_-\rangle)^\dagger$.
This analysis is completely consistent with the results in Eqs. \ref{eq-WDIII} and \ref{eq-Wnu}.

(3) To understand the origin of the MFBs localized at the boundary of the strip,
we consider another integer topological invariant defined as (see Fig.
\ref{fig-fig4}b)\cite{satofb},
\begin{equation}
\mathcal{W}(\textbf{k}_{\parallel}) = {1 \over 2\pi} \Im \int d \textbf{k}_{\perp} \partial \textbf{k}_{\perp} \ln \text{Det}Q (\textbf{k}) \in \mathbb{Z},
\label{eq-Wp}
\end{equation}
where $\textbf{k}_{\perp}$ is the direction {\it perpendicular} to the strip. If this number is nonzero, it means topological protected edge
states with zero energy can be found at the two boundaries. The number of edge states is defined by $|\mathcal{W}(\textbf{k}_{\parallel})|$.

{\bf Mirror symmetries and mirror winding numbers} For the $T_{xy}$ model, $M_{1,2}^x = -
\epsilon_k^x \sigma_z\otimes \rho_0 - \Gamma \sigma_z\otimes \rho_z \ - ( k_x \alpha'
\pm t) \sigma_z\otimes \rho_y - \Delta\sigma_y\otimes \rho_y$, and $M_{1,2}^y = -\epsilon_k^y
\sigma_z\otimes \rho_0 - \Gamma \sigma_z\otimes \rho_z + ( k_y
\alpha\pm t) \sigma_0\otimes \rho_x - \Delta\sigma_y\otimes \rho_y$. Here $\epsilon_k^{x,y}=k_{x,y}^2/2m-\mu$, and $\rho$
and $\sigma$ are Pauli matrices which do not have the meaning defined in the main text. These two models belong
to AIII model with only
chiral symmetry and the topological invariant is characterized by $\mathbb{Z}$ in 1D.

For the $T_{x0}$ model only the mirror spaces generated by
$\mathcal{M}_y$ is important. For the case studied in the main text,
$t_{\uparrow}=t_{\downarrow} = t$, $\alpha=\alpha'$ and
$M_{1,2}^y=-(\epsilon_k^y \pm t)\sigma_z\otimes \rho_0 + \Gamma
\sigma_z\otimes \rho_z - k_x \alpha' \sigma_z\otimes \rho_y - \Delta
\sigma_y\otimes \rho_y$. These two Hamiltonians belong to
topological BDI class with TR, PH and chiral symmetries, and the
invariant is characterized by $\mathbb{Z}$ in 1D.

Since all the above mirror subspaces have chiral symmetry (AIII for $T_{xy}$ and BDI for
$T_{x0}$), we assume $s_{\alpha}$ is the chiral symmetry operator for
$M_{\alpha}$ (see Eq. \ref{eq-My} and \ref{eq-Mx}), where
\begin{equation}
        u_{\alpha} s_{\alpha} u_{\alpha}^{-1} =
\begin{pmatrix}
        \bf{I}_2  & 0  \\
           0  & -\bf{I}_2
\end{pmatrix},
        u_{\alpha} M^{\alpha}_i u_{\alpha}^{-1} =
        \begin{pmatrix}
                        0  & q^{\alpha}_i  \\
       q_i^{\alpha\dagger}  & 0
        \end{pmatrix}.
\end{equation}
Then the mirror winding number in these mirror symmetric invariant
subspaces can be calculated using Eq. \ref{eq-Wp}, in which
$\mathcal{H}$ and $\mathbb{S}$ are replaced by $M_i^{\alpha}$ and
$s_\alpha$. For $T_{xy}$ model, $s_x=\sigma_y\otimes\sigma_z$ and
$s_y=\sigma_x\otimes\sigma_0$; and for $T_{x0}$ model $s_y=\sigma_x\otimes\sigma_0$.

There is an important difference between these two models ($T_{xy}$ and $T_{x0}$) thought the
definition of mirror winding numbers
is exactly the same. In $T_{xy}$ model, the two mirror subspaces
$M_1^{\alpha}$ and $M_2^{\alpha}$, which belongs to AIII class in
both $k_x$ and $k_y$ directions, are not independent, but are
related with each other by TR and PH symmetries, thus there is just
only one independent winding number $\mathcal{W}^{\alpha}$ along each
direction. In the main text this number is calculated
using $M_1^{\alpha}$ since
$\mathcal{W}^{\alpha}=\mathcal{W}_1^{\alpha} \equiv
\mathcal{W}_2^{\alpha}$. However in $T_{x0}$ model the two mirror
symmetric invariant subspaces belong to BDI class when along $k_y$
direction, which have independent TR and PH symmetries, thus the
mirror winding numbers are independent. For this reason the mirror
topological properties in BDI model are labeled by two integer
numbers $\mathcal{W}^y_{1}\oplus \mathcal{W}^y_{2}$,  where
$\mathcal{W}^y_{1,2}$ are calculated for $M_{1,2}^{y}$ respectively.
We do not have to consider the invariant along $k_x$ direction,
which is trivial.

Below we summarize the results for mirror winding numbers for all
phases in the phase diagrams in the main text. For $T_{xy}$ model,
the topological gapped phase has $\mathcal{W}^{x}_1=1$ and
$\mathcal{W}^{y}_1=-1$; the phase with 4 Dirac cones has
$\mathcal{W}^{x}_1=1$ and $\mathcal{W}^{y}_1=0$; and the regime with
8 Dirac cones has $\mathcal{W}^{x}_1=0$ and $\mathcal{W}^{y}_1=0$.
For $T_{x0}$ model, the right phase with 2 Dirac cones has
$\mathcal{W}^y_{1}\oplus \mathcal{W}^y_{2}=0\oplus-1$; the left
phase with 2 Dirac cones has $\mathcal{W}^y_{1}\oplus
\mathcal{W}^y_{2}=-1\oplus0$; the regime with 4 Dirac cones has
$\mathcal{W}^y_{1}\oplus \mathcal{W}^y_{2}=-1\oplus -1$; the regime
with 6 Dirac cones has $\mathcal{W}^y_{1}\oplus
\mathcal{W}^y_{2}=-1\oplus0$ and the fully gapped topological phase
has $\mathcal{W}^y_{1}\oplus \mathcal{W}^y_{2}=-1\oplus -1$.

{\bf Effect of interlayer pairings}. For convenience we use the
following notations to define all the possible scatterings
between the two layers with $g_{\alpha\beta\gamma\delta}$ being
the scattering strength and the Greek alphabet being the layer index,
\begin{eqnarray}
             F_{\alpha\beta\gamma\delta} : &&= g_{\alpha\beta\gamma\delta}
             \sum_{{\bf k,q}}  (c_{\alpha\textbf{k}\uparrow}^\dagger c_{\beta-\textbf{k}\downarrow}^\dagger)
             (c_{\gamma\textbf{q}\downarrow} c_{\delta-\textbf{q}\uparrow})  \\
             \nonumber
             && \simeq {g_{\alpha\beta\gamma\delta} \over g} \cdot [\sum_\textbf{k}
             \Delta_{\alpha\beta}^* c_{\delta \textbf{k}\uparrow} c_{\gamma-\textbf{k}\downarrow} + \text{h.c} + {\Delta_{\alpha\beta}^* \Delta_{\delta\gamma} \over g}].
             \end{eqnarray}
We define the intralayer and interlayer pairings using the
following way: $\Delta_{\alpha\beta} = g \langle \sum_{\textbf{k}}
c_{\alpha \textbf{k}\uparrow} c_{\beta -\textbf{k}\downarrow}
\rangle$, and $\Delta_{\alpha\beta}^* = g \langle \sum_{\textbf{k}}
c_{\beta -\textbf{k}\downarrow} c_{\alpha \textbf{k}\uparrow}
\rangle$. The leading pairings in the form $\Delta_{\alpha\alpha} =
\Delta_{\alpha}$ have been used in the main text. We have to
sum up all possible Feynman diagrams to construct the mean-field
BdG equation. To this end, we define $g_{1111} = g_{2222} = g$,
$g_{1112} = g_{1121} = \cdots = g'$, and $g_{1122} = g_{1221} =
\cdots = g''$, with $g'' < g' \ll g$. Then we can write
\begin{equation}
U= U^\dagger +U^- + U_0,
\end{equation}
where $U^\dagger = \sum_\textbf{k} \Delta_1 c_{1\textbf{k}\uparrow}^\dagger c_{1-\textbf{k}\downarrow}^\dagger +
\Delta_2 c_{2\textbf{k}\uparrow}^\dagger c_{2-\textbf{k}\downarrow}^\dagger +  \Delta_3 c_{1\textbf{k}\uparrow}^\dagger c_{2-\textbf{k}\downarrow}^\dagger +
\Delta_4 c_{2\textbf{k}\uparrow}^\dagger c_{1-\textbf{k}\downarrow}^\dagger$, $U^- = (U^\dagger)^\dagger$, and
$U_0 = -\sum_{\alpha,\beta,\gamma,\delta = 1, 2} {g_{\alpha\beta\gamma\delta} \over g^2} \Delta_{\alpha\beta}^* \Delta_{\delta\gamma}$.
Here we have defined four order parameters due to all possible scatterings. In
this condition, for the BdG equation in Eq. \ref{eq-H}, the off-diagonal block
matrix $\bar{\Delta}$ have to be replaced by
\begin{equation}
\bar{\Delta} = \begin{pmatrix}
0                & \Delta_1             & 0             & \Delta_3 \\
-\Delta_1        & 0                    & \Delta_4      & 0         \\
0                & -\Delta_4            & 0             & \Delta_2   \\
-\Delta_3        & 0                    & -\Delta_2     &  0
\end{pmatrix},
\label{eq-Delta}
\end{equation}
where $\Delta_1 = - \Delta_{11} - {g' \over g} (\Delta_{12} + \Delta_{21}) - {g'' \over g} \Delta_{22}$,
$\Delta_2 = - \Delta_{22} - {g' \over g} (\Delta_{12} + \Delta_{21}) - {g'' \over g} \Delta_{11}$,
$\Delta_3 = - {g' \over g} (\Delta_{11} + \Delta_{22}) - {g'' \over g} (\Delta_{12} + \Delta_{21})$,
$\Delta_4 = - {g' \over g} (\Delta_{11} + \Delta_{22}) - {g'' \over g} (\Delta_{21} + \Delta_{12})$.
The TR symmetry for this new order parameter is still determined by the operator $\mathbb{T}$ used in the
main text, but now, $\Delta_{1,2}$ and $\Delta_{3,4}$ should respect some restrict relations.
If we define $\mathcal{O} = \Delta_{12} + \Delta_{21}$, then we can simplify the condensation energy
\begin{eqnarray}
-U_0 =&& {1\over g} (|\Delta_{11}|^2+|\Delta_{22}|^2) + {g' \over g^2} \mathcal{O} (\Delta_{11} + \Delta_{22} + \Delta_{11}^* + \Delta_{22}^*)  \nonumber \\
     && + {g'' \over g^2}  (\mathcal{O}^2 + \Delta_{11} \Delta_{22}^* + \Delta_{11}^* \Delta_{22}).
\end{eqnarray}
This energy should be used to replace the last two terms in Eq. \ref{eq-Omega}.
The self-consistent numerical sumulation demonstrate that all the conclusions obtained in the main
text are still valid in the presence of these interlayer pairings.

{\bf Acknowledgements} We thank Y. X. Zhao, F. Zhang and Z. D. Wang for valuable discussions. 
C.C., J.L. and M.G. are supported by Hong Kong RGC/GRF Projects (No. 401011, 401113) and The
Chinese University of Hong Kong Focused Investments Scheme.
B.H. is supported by Natural Science Foundation of Jiangsu Province
under Grant No. BK20130424. C.Z. is supported by 
ARO (W911NF-12-1-0334)  AFOSR (FA9550-13-1-0045) and NSF (PHY-1505496).

{\textbf{Author Contributions}} B.H. and C.C contributes equally to
this work. M.G. conceived the idea and supervised the project. B.H.
and C.C. performed the self-consistent calculation and symmetry
analysis. M.G. and C.Z. wrote this manuscript. All authors
participated in discussions about this work. Correspondence and
requests for materials should be addressed to M.G.
(skylark.gong@gmail.com) and C.Z. (chuanweizhang@utdallas.com).

{\textbf{Competing Interests}}  The authors declare no competing financial interests.

\end{document}